# SpiDy.jl: open-source Julia package for the study of non-Markovian stochastic dynamics

Stefano Scali [1,¶], Simon Horsley [1], Janet Anders [1,2], and Federico Cerisola [1,3]

**1** Department of Physics and Astronomy, University of Exeter, Exeter EX4 4QL, United Kingdom **2** Institute of Physics and Astronomy, University of Potsdam, 14476 Potsdam, Germany **3** Department of Engineering Science, University of Oxford, Oxford OX1 3PJ, United Kingdom. ¶ Corresponding author

## Summary

SpiDy.jl solves the non-Markovian stochastic dynamics of interacting classical spin vectors and harmonic oscillator networks in contact with a dissipative environment. The methods implemented allow the user to include arbitrary memory effects and colored quantum noise spectra. In this way, SpiDy.jl provides key tools for the simulation of classical and quantum open systems including non-Markovian effects and arbitrarily strong coupling to the environment. Among the wide range of applications, some examples range from atomistic spin dynamics to ultrafast magnetism and the study of anisotropic materials. We provide the user with Julia notebooks to guide them through the various mathematical methods and help them quickly set up complex simulations.

## Statement of need

The problem of simulating the dynamics of interacting rotating bodies and harmonic oscillator networks in the presence of a dissipative environment can find a vast range of applications in the modeling of physical systems. This task is rendered particularly challenging when one desires to capture the non-Markovian effects that arise in the dynamics due to strong coupling with the environment. SpiDy.jl is a library that allows the user to efficiently simulate these systems to obtain both detailed dynamics and steady state properties.

A relevant example of the applicability of SpiDy.jl is the modeling of spins at low temperatures and at short timescales, which is a fundamental task to address many open questions in the field of magnetism and magnetic material modeling (Halilov et al., 1998). State-of-the-art tools such as those developed for atomistic spin dynamics simulations are based on solving the Landau–Lifshitz–Gilbert (LLG) equation (Evans et al., 2014). Despite their massive success, these tools run into shortcomings in accurately modeling systems at low temperatures and for short timescales where environment memory effects have been observed (Ciornei et al., 2011; Neeraj et al., 2020). Recent work has focused on developing a comprehensive quantum-thermodynamically consistent framework suitable to model the dynamics of spins in magnetic materials while addressing these shortcomings (Anders et al., 2022). This framework includes strong coupling effects to the environment such as non-Markovian memory, colored noise, and quantum-like fluctuations. At its core, SpiDy.jl implements the theoretical framework introduced in Anders et al. (2022), allowing for the study of environment memory effects and anisotropic system-environment coupling. SpiDy.jl can be readily adopted for atomistic spin dynamics simulations (Barker & Bauer, 2019; Evans et al., 2014), ultrafast magnetism (Beaurepaire et al., 1996), and ferromagnetic and semiconductive systems exhibiting anisotropic damping (Chen et al., 2018). A further set of applications stems from the extension of SpiDy.jl

to handle the non-Markovian stochastic dynamics of harmonic oscillators. This model will be of interest in the field of quantum thermodynamics where harmonic oscillators play a key role in modeling open quantum systems. The package is written in pure Julia and we take advantage of the efficient DifferentialEquations.jl package (Rackauckas & Nie, 2017) by reducing evaluation redundancy, using callbacks, and pre-allocations.

The software package has seen a wide range of applications to date. Firstly, the convenience of three independent environments in SpiDy.jl finds application in the microscopic modeling of spins affected by noise due to vibrations of the material lattice (Anders et al., 2022). SpiDy.jl also found application in the demonstration of the quantum-to-classical correspondence at all coupling strengths between a spin and an external environment (Cerisola et al., 2024). Here, the temperature dependence of the spin steady-state magnetization obtained with SpiDy.jl is successfully compared with the classical mean force state of the system. Hartmann et al. (2023) take advantage of the customizable coupling tensor in SpiDy.jl to explore the anisotropic effects of the environment on the system. Berritta et al. (2023) use SpiDy.jl as a sub-routine to build quantum-improved atomistic spin dynamics simulations. In the paper, the authors take advantage of the customizable power spectrum to implement ad-hoc simulations matching known experimental results. Lastly, with an eye to the harmonic oscillator side, SpiDy.jl is used to match the quantum harmonic oscillator dynamics with its stochastic counterpart (Hogg et al., 2024). Here, the authors exploit the recent implementation of harmonic oscillator dynamics.

## Overview

To model a system of interacting classical spin vectors, SpiDy.jl solves the generalized stochastic LLG equation (Anders et al., 2022)

$$\frac{d\mathbf{S}_n(t)}{dt} = \frac{1}{2}\mathbf{S}_n(t) \times \left[\sum_{m \neq n} J_{n,m}\mathbf{S}_m(t) + \mathbf{B} + \mathbf{b}_n(t) + \int_{t_0}^{t} dt' K_n(t-t')\mathbf{S}_n(t')\right], \quad (1)$$

where $\mathbf{S}_n(t)$ represents the $n$-th spin vector, the interaction matrix $J_{n,m}$ sets the interaction strength between the $n$-th and $m$-th spins, $\mathbf{B}$ is the external field, which determines the natural precession direction and frequency of the spins in the absence of interaction, and $\mathbf{b}_n(t)$ is the time-dependent stochastic field induced by the environment. Finally, the last integral term in Eq.(1) gives the spin dissipation due to the environment, including non-Markovian effects accounted for by the memory kernel matrix $K_n(t)$. In addition, SpiDy.jl also allows to study the stochastic dynamics of coupled harmonic oscillator networks in a form similar to Eq.(1).

In conclusion, SpiDy.jl implements the stochastic dynamics of coupled integro-differential equations to model systems of interacting spins or harmonic oscillator networks subject to environment noise. Among others, some of the key features of the package include:

- Coloured stochastic noise that satisfies the FDR and accounts for both classical and quantum bath statistics.
- Simulation of non-Markovian system dynamics due to the memory kernel $K_n(t)$.
- Custom system-environment coupling tensors, allowing for isotropic or anisotropic couplings. Both amplitudes and geometry of the coupling can be specified.
- Choice between local environments, i.e. distinct baths acting on the single sub-system, or a single common environment.

We show an example run of the stochastic trajectories in the case of a single spin in Figure 1. The code used to generate this example can be found in the GitHub repository of SpiDy.jl.

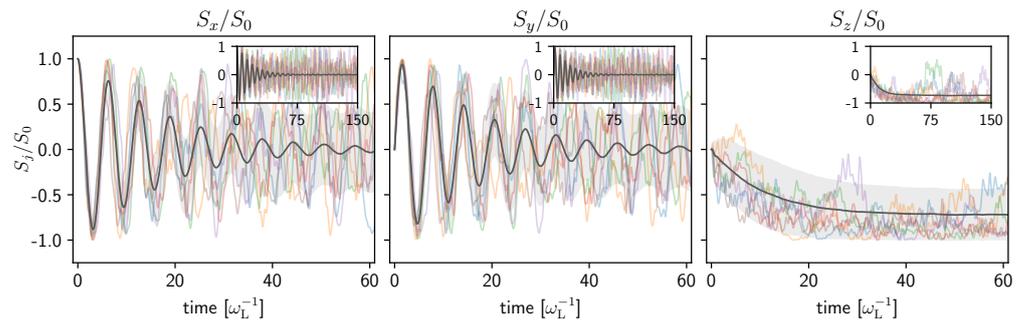

Figure 1: **Single-spin dynamics.** Dynamics of the $x$, $y$, and $z$ spin components. The components are normalized against the total spin length $S_0$ and time axes are expressed in units of the Larmor frequency $\omega_L$ ($\omega_L = |\mathbf{B}|$ in Eq.(1)). We show an example set of 5 stochastic trajectories of the spin dynamics (colored semi-transparent lines) together with their stochastic average (gray solid line). Note that, while we show only 5 trajectories for clarity, the average dynamics is obtained from 10000 trajectories. We also represent the range of one standard deviation from the average dynamics (gray-shaded area). In the inset, we show the convergence of the same dynamics towards the steady state at longer times. This example is obtained using the Lorentzian parameters "set 1" obtained from Anders et al. (2022). The code used to generate the stochastic trajectories is included in the GitHub repository.

# Acknowledgements

SS and FC thank Marco Berritta and Charlie Hogg for insightful suggestions for the implementation of the spin-spin coupling and the harmonic oscillator dynamics. SARH and JA thank the Royal Society for their support. SS is supported by a DTP grant from EPSRC (EP/R513210/1). SARH acknowledges the Royal Society and TATA for financial support through the Grant URFR 211033. JA and FC acknowledge funding from EPSRC (EP/R045577/1). FC gratefully acknowledges funding from the Foundational Questions Institute Fund (FQXi–IAF19-01).

# References

Anders, J., Sait, C. R. J., & Horsley, S. A. R. (2022). Quantum brownian motion for magnets. *New Journal of Physics*, *24*(3), 033020. https://doi.org/10.1088/1367-2630/ac4ef2

Barker, J., & Bauer, G. E. W. (2019). Semiquantum thermodynamics of complex ferrimagnets. *Physical Review B*, *100*(14). https://doi.org/10.1103/physrevb.100.140401

Beaurepaire, E., Merle, J.-C., Daunois, A., & Bigot, J.-Y. (1996). Ultrafast spin dynamics in ferromagnetic nickel. *Physical Review Letters*, *76*(22), 4250–4253. https://doi.org/10.1103/physrevlett.76.4250

Berritta, M., Scali, S., Cerisola, F., & Anders, J. (2023). Accounting for quantum effects in atomistic spin dynamics. *arXiv Preprint*. https://doi.org/10.48550/arXiv.2305.17082

Cerisola, F., Berritta, M., Scali, S., Horsley, S. A. R., Cresser, J. D., & Anders, J. (2024). Quantum-classical correspondence in spin-boson equilibrium states at arbitrary coupling. *New Journal of Physics*. https://doi.org/10.1088/1367-2630/ad4818

Chen, L., Mankovsky, S., Wimmer, S., Schoen, M. A. W., Körner, H. S., Kronseder, M., Schuh, D., Bougeard, D., Ebert, H., Weiss, D., & Back, C. H. (2018). Emergence of anisotropic gilbert damping in ultrathin fe layers on GaAs(001). *Nature Physics*, *14*(5), 490–494. https://doi.org/10.1038/s41567-018-0053-8

Ciornei, M.-C., Rubí, J. M., & Wegrowe, J.-E. (2011). Magnetization dynamics in the inertial regime: Nutation predicted at short time scales. *Physical Review B*, *83*(2).

https://doi.org/10.1103/physrevb.83.020410

Evans, R. F. L., Fan, W. J., Chureemart, P., Ostler, T. A., Ellis, M. O. A., & Chantrell, R. W. (2014). Atomistic spin model simulations of magnetic nanomaterials. *Journal of Physics: Condensed Matter*, *26*(10), 103202. https://doi.org/10.1088/0953-8984/26/10/103202

Halilov, S. V., Eschrig, H., Perlov, A. Y., & Oppeneer, P. M. (1998). Adiabatic spin dynamics from spin-density-functional theory: Application to fe, co, and ni. *Physical Review B*, *58*(1), 293–302. https://doi.org/10.1103/physrevb.58.293

Hartmann, F., Scali, S., & Anders, J. (2023). Anisotropic signatures in the spin-boson model. *arXiv Preprint*. https://doi.org/10.48550/arXiv.2305.16964

Hogg, C., Glatthard, J., Cerisola, F., & Anders, J. (2024). Stochastic simulation of dissipative quantum oscillators. In *In preparation*.

Neeraj, K., Awari, N., Kovalev, S., Polley, D., Hagström, N. Z., Arekapudi, S. S. P. K., Semisalova, A., Lenz, K., Green, B., Deinert, J.-C., Ilyakov, I., Chen, M., Bawatna, M., Scalera, V., d'Aquino, M., Serpico, C., Hellwig, O., Wegrowe, J.-E., Gensch, M., & Bonetti, S. (2020). Inertial spin dynamics in ferromagnets. *Nature Physics*, *17*(2), 245–250. https://doi.org/10.1038/s41567-020-01040-y

Rackauckas, C., & Nie, Q. (2017). DifferentialEquations.jl – a performant and feature-rich ecosystem for solving differential equations in julia. *The Journal of Open Research Software*, *5*(1). https://doi.org/10.5334/jors.151